%% file: Main.tex
\documentclass[prodmode]{acmsmall} 

\usepackage[ruled]{algorithm2e}

\SetAlFnt{\small}
\SetAlCapFnt{\small}
\SetAlCapNameFnt{\small}
\SetAlCapHSkip{0pt}
\IncMargin{-\parindent}

\usepackage{array}
\newcolumntype{L}[1]{>{\raggedright\let\newline\\\arraybackslash\hspace{0pt}}m{#1}}

\usepackage{textcomp} 

\usepackage{graphicx}

\setcopyright{rightsretained}

\doi{0000001.0000001}

\issn{1234-56789}

\begin{document}

\markboth{}{Appraising Human Impact on Watersheds}

\title{Appraising Human Impact on Watersheds: The Feasibility of Training Citizen Scientists to make Qualitative Judgments}
\author{
Alina Striner
\affil{University of Maryland, College Park}
Jennifer Preece
\affil{University of Maryland, College Park}
}

\begin{abstract}
Citizen science often requires volunteers to perform low-skill tasks such as counting and documenting environmental features. In this work, we contend that these tasks do not adequately meet the needs of citizen scientists motivated by scientific learning. We propose to provide intrinsic motivation by asking them to notice, compare, and synthesize qualitative observations. We describe the process of learning and performing qualitative assessments in the domain of water quality monitoring, which appraises the impact of land use on habitat quality and biological diversity.  We use the example of water monitoring because qualitative watershed assessments are exclusively performed by professionals, yet do not require specialized tools, making it an excellent fit for volunteers. Within this domain, we observe and report on differences in background and training between professional and volunteer monitors, using these experiences to synthesize findings about volunteer training needs. Our findings reveal that to successfully make qualitative stream assessments, volunteers need to: (1) experience a diverse range of streams, (2)  discuss judgments with other monitors, and (3)  construct internal narratives about water quality.  We use our findings to describe how different technologies may support these needs and generalize our findings to the larger citizen science community.
\end{abstract}

%
%


%
%

\terms{Citizen Science, Qualitative Assessment, Water Quality, Stream Monitoring}

\keywords{Citizen Science, Expert Interviews, Qualitative Assessment Task, Water Quality Monitoring}


\begin{bottomstuff}
Author's addresses: 
\end{bottomstuff}

\maketitle

 \input{Introduction}

\input{Bg_QualitativeJudgments}
\input{Bg_StreamMonitoring}

\input{Method}

\section{Results}
\input{Results_Process}

\input{Results_Themes}


\input{Discussion}

 \input{Conclusion}

\appendixhead{ZHOU}

\begin{acks}
\end{acks}

\bibliographystyle{ACM-Reference-Format-Journals}
\bibliography{Ref}


\medskip

\end{document}

%% file: Introduction.tex
\section{Introduction}

Citizen science is a form of crowdsourcing that allows volunteers to collaborate with researchers on scientific data collection and analysis~\cite{bonney2009citizen}. A long-term goal of environmental citizen science research is to engage local communities in ``democratic ownership'' of environmental projects to provoke large-scale habitat accountability and civic protection efforts~\cite{riesch2013combining}.  

Literature suggests that many citizen scientists are engaged and motivated by science-based learning and discovery~\cite{raddick2009galaxy}, but high training costs and limited resources often result in volunteers participating in only unskilled work. While practical, tasking volunteers with unskilled work can lead to complications, including boredom and disengagement~\cite{flow1990psychology,kiili2014flow}, which may decrease data quality and participant retention~\cite{eberbach2009everyday}.  Citizen science work is often comprised of unskilled tasks such as documenting, counting, identifying, classifying, and transcribing environment features~\cite{bowser2016cooperative,wiggins2013free,cohn2008citizen,raddick2009galaxy,CrowstonGravitySpy2017,eveleigh2014designing}. Instead of providing scientific education or increasing the required skill, projects may engage community members through gamification~\cite{bowser2013using} and novelty~\cite{JacksonNoveltyPopUp}. 

Engaging citizen scientists in substantive, high-level tasks has potential to improve data quality and participant retention~\cite{striner2016streambed,WigginsCrowston_DataQuality2011}. This paper proposes engaging those citizen scientists motivated by learning goals with qualitative assessments---high-level assessment tasks focused on naturalistic, inductive interpretations of an environment~\cite{patton2002qualitative,denzin2011sage}. A traditional barrier to assigning high-skill tasks to volunteers is that developing the required skills is difficult and time consuming. Qualitative assessments are usually made by professionals who learn through iterative personal experiences.  Learning to make qualitative assessments requires access to a variety of heterogeneous environments, which is not practical for volunteers. 

Recent advances in immersive technologies, {360\textdegree} panoramic videos, augmented reality (AR), and virtual reality (VR) have effectively been able to recreate physical environments necessary for training with minimal cost~\cite{cummings2016immersive,ahn2014short}. These virtual environments allow learners to experience several diverse environments in a short period of time. Our research thus considers the feasibility of training citizen scientists to make qualitative assessments using these technologies.

We situate our research in water monitoring for four reasons: (1)  a standard protocol is used by many monitoring groups~\cite{Pond:Personal}, (2) assessments are exclusively made by professionals, (3) assessment does not require specialized tools, making it an ideal task for volunteers, and (4) assessment relies on first-hand experiences at stream sites, which immersive technology has the power to recreate. We consider the need for and feasibility of training citizen scientists to make qualitative assessments of streams and watersheds using a Rapid Bioassessment Protocol (RBP)~\cite{barbour1999rapid} which was developed by the Environmental Protection Agency (EPA) and is used nationally.

This work proposes training volunteers to make qualitative assessments, which intrinsically motivate investigators by asking them to notice, compare, and synthesize their observations.
To accomplish this objective, we observe and report on differences in background and training between professional and volunteer monitors, using these experiences to synthesize findings about volunteer training needs. Our findings reveal that to successfully make qualitative stream assessments, volunteers need to: (1) experience a diverse range of streams, (2) discuss judgments with other monitors, and (3) construct internal narratives about water quality. Using these findings, we discuss how {360\textdegree} videos, AR, and VR may be able to support this qualitative learning task, and generalize our findings to the broader citizen science community.  

In summary, our contribution: (1) proposes training citizen scientists to make qualitative assessments to motivate volunteers, (2) describes how experts learn to make qualitative assessments of streams, and compares their learning process to volunteer experiences, and (3) describes how {360\textdegree} videos, AR, and VR may help support qualitative assessment training. 

%% file: Bg_QualitativeJudgments.tex
\section{Background}
We provide our operating definition of qualitative judgments and an overview of how they are learned and performed, with specific attention to water monitoring and the EPA Rapid Bioassessment Protocol.

\subsection{Qualitative Judgments}
Qualitative judgments are intuitive representations of mental models that scaffold the way people speak, listen, and interpret the world around them.  In this paper, we define making qualitative judgments as using intuitive heuristics~\cite{gilovich2002heuristics} to discern relationships between latent variables. Researchers often use these contextual judgments to describe and evaluate knowledge that is difficult to teach and evaluate procedurally~\cite{perreault1989_QualitativeJudgmentNominal}. 

Making qualitative judgments requires evaluators to scaffold knowledge into internal cognitive maps, which chunk interconnected ideas into concepts and models that help them notice, compare, and identify patterns~\cite{warren2005teaching}. Over time, this scaffold is shaped by experiential learning~\cite{kerfoot2003learning,hmelo2007fish}, cyclically fashioned together from multimodal and multisensory information~\cite{haverkamp2001application}, and from discussion, which is interpreted, and interwoven into previous experience~\cite{Young:2016}. 

Making qualitative judgments is difficult and often reserved for professionals.  For instance, clinical psychiatrists recommend medical treatments based on qualitative assessments of patient conditions~\cite{pomerantz2016_QualitativePsychology}, and neurologists use qualitative assessments to characterize patient pain~\cite{morse2015_QualitativePain}. Similarly, trainers use qualitative methods to evaluate athlete performance~\cite{knudson2013_Sportqualitative}, and voice teachers use qualitative pedagogical techniques to diagnose vocal problems~\cite{mckinney1994_QualitativeSingingdiagnosis}. 

Scientists likewise use qualitative heuristics to understand nature. For instance, climate scientists use qualitative ``climate proxies,'' like perennial cherry blossom bloom periods, to assess the cumulative effect of climate change on changes in temperature or rainfall~\cite{striner_borovikov_2017}. Likewise, paleoecologists use fossil and sediment proxies to interpret and reconstruct ecosystems and environmental conditions of the past~\cite{birks2011_Qualitative_PaleoClimate}. On a larger scale, astronomers pair qualitative heuristics with quantitative analyses to understand the conditions of unusual cosmic phenomena. For instance, astronomers draw conclusions about supernovae composition using heuristic interpretations of Hubble telescope images of scattered light echoes~\cite{rest2008_Qualitative_LightEchoes}. 

\subsection{Learning to Make Qualitative Judgments}
In the natural sciences, learning to make qualitative judgments consists of (1) \textit{noticing} and \textit{observing} phenomena, (2) \textit{comparing} observations to expectations, (3) \textit{synthesizing} observations into patterns, (4) \textit{scaffolding} patterns into a cognitive map~\cite{eberbach2009everyday}, and (5) \textit{updating} the cognitive map using new information. First, learners notice phenomena,  and make observations; this requires them to know when to ask questions, and what question to ask (\textit{e.g.}``what thing am I looking at?''). Learners then synthesize patterns by comparing observations to internal webs of information~\cite{alberdi2000accommodating}, and recognizing similarities and differences~\cite{crossan1999organizational}. Finally, learners scaffold observations into an internal cognitive map, and iteratively update the map by making new observations~\cite{kerfoot2003learning}. Learners also update their cognitive map through discussion, which helps them form shared interpretation of meaning through metaphors, which allow them to transfer information between familiar and new domains ~\cite{crossan1999organizational}. 

Teaching qualitative judgments is often challenging because experiences are subjective and difficult to surface, examine, compare, and explain~\cite{crossan1999organizational}. In the examples above, professionals learn to make assessments through experiential learning; they make observations and identify patterns by iteratively getting feedback from patients and students. Unlike professionals who learn from feedback loops, researchers learn to make intuitive judgments by studying related quantitative and qualitative data; for instance, in the astronomy example above, quantitative data about how light scatters helps astronomers visually interpret low resolution telescope images. 

Professionals are able to make qualitative judgments because they have more experience---however, with enough experience, amateurs, too, are able to learn to make effective qualitative assessments. For instance, amateur bakers often use qualitative assessments to `troubleshoot'' finicky recipes~\cite{Mimi_Macarons}. Similarly, successful amateur investors often use intuitive heuristics to "predict" company valuations and stock market changes~\cite{roberts1959_IntuitiveStockMarket}.

%% file: Bg_StreamMonitoring.tex
\subsection{Qualitative Judgments in Stream Monitoring}
Stream ecology groups perform qualitative stream monitoring and assessment to report on the balance between development and environmental protection. For instance, the Maryland Department of Natural Resources uses monitoring to benchmark long term stream quality trends, examine impact of land use on habitat quality and biological diversity, and assess cumulative impacts to streams~\cite{roth2002biological}. Ecology groups also use data from monitoring projects to advocate for land use that protects regional watersheds~\cite{barbour1999rapid}.  

To effectively protect watersheds, Roth and Davis~\cite{roth2002biological} describe the need for cost-effective solutions that inform the public about stream conditions and reduce sampling program costs. For this reason, training citizen scientist monitors may not only increase volunteer retention, but may also benefit water monitoring agencies by creating a free task force and engaging the public in water quality issues.
\subsection{Using the EPA's Rapid Bioassessment Protocol (RBP)}
Although many watershed monitoring protocols exist, this work focuses on the EPA's Rapid Bioassessment Protocol (RBP) \cite{barbour1999rapid}, a qualitative metric of stream conditions. The RBP is part of a larger monitoring process that includes several quantitative assessment measures. These include measuring PH and temperature, and counting fish and benthic macroinvertebrates, small animals that live among stones, sediments and aquatic plants, whose presence is indicative of stream quality~\cite{barbour1999rapid}.  We focus on this protocol because it used by many monitoring groups~\cite{Pond:Personal}.

The RBP protocol is made up of 13 measures, described below. Each metric is assessed on a 10 or 20 point scale. An example metric, channel alteration, is shown in figure~\ref{fig:ChannelAlteration}. 
\begin{enumerate}
\item \textit{Epifaunal substrate} evaluates opportunity for insect colonization and fish cover
\item \textit{Embeddedness} of cobble and boulders in stream sediment appraises the surface area available to fish and macroinvertebrates for shelter and spawning 
\item \textit{Pool substrate characterization} evaluates the mixture of stream bottom substrates
\item \textit{Velocity depth combinations} notes diversity of water velocity and depth patterns
\item \textit{Variability of pool environments} characterizes the diversity of stream pools
\item \textit{Sediment deposition} estimates sediment accumulation at the bottom of a stream
\item \textit{Channel flow status} describes the degree to which a channel is filled with water
\item \textit{Channel alteration} estimates the extent to which a steam’s shape has been altered
\item \textit{Frequency of riffles and bends}, judges the heterogeneity of a stream's shape
\item \textit{Channel sinuosity} evaluates the curvature of the stream
\item \textit{Bank stability} considers the extent to which banks have eroded
\item \textit{Bank vegetation protection} values the quality of vegetation protecting the stream
\item \textit{Riparian zone width} delineates the vegetative buffer between a stream bank and runoff pollutants
\end{enumerate}

%% file: Method.tex
\section{Method}
The goal of this work is to assess the viability of training citizen science volunteers to make qualitative assessments of streams and watersheds using the EPA’s Rapid Bioassessment Protocol (RBP)~\cite{barbour1999rapid}. To compare current teaching methods employed by professionals and citizen science groups, the first author (1) observed and participated in RBP training and data collection with professional water monitors and with volunteers, (2) informally discussed water monitoring methods with ecologists, and (3) conducted semi-structured interviews with professional monitors.
\subsection{Training and Data Collection}
\subsubsection{Professional RBP Training and Data Collection Process}
The first author participated and observed water quality training with 4 professional water monitoring groups that included between 2 and 6 monitors. Each session lasted approximately 3 hours. Data collection took place at either 2 or 3 100-meter sites at different points of a stream.

In the larger teams, RBP assessment was paired with quantitative monitoring tasks. First, monitors measured the depth and velocity of the stream at different points and measured quantitative measures such as temperature and PH. Then, they noted the absence or presence of stream characteristics like bedrock, and clay, and tallied the number of woody debris and roots in and around the stream. In addition, they collected samples of fish and macroinvertebrates.

\begin{figure}[th]
\centering
  \includegraphics[width=1\columnwidth]{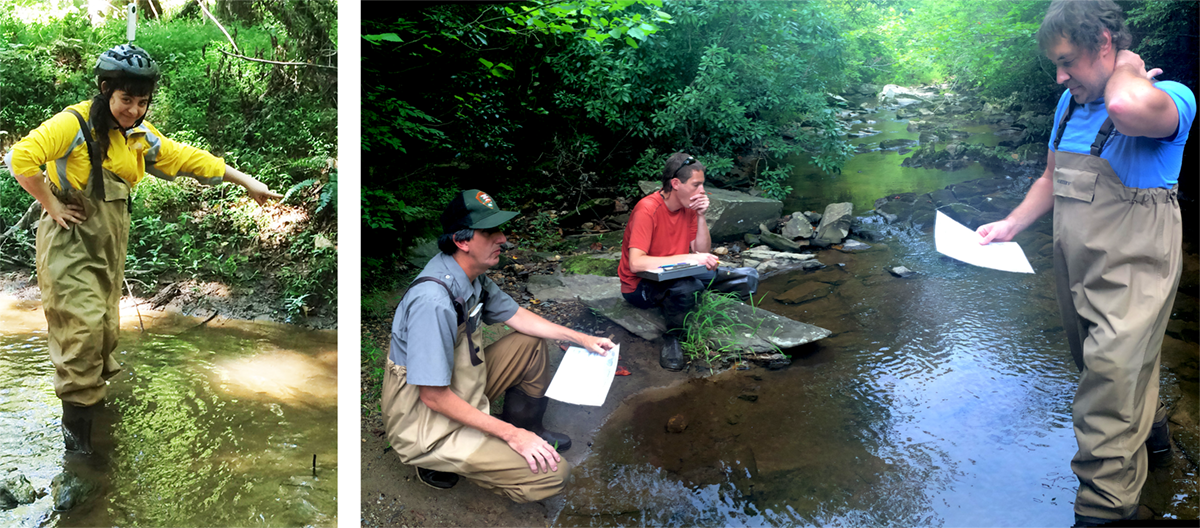}\caption{left: A photo of the first author learning to make assessments. Right: A group of water biologists discussing their observations and EPA protocol assessments.}~\label{fig:ExpertAssessment}
\end{figure}

During each session, the first author observed how monitors made RBP qualitative assessments. Monitors first made personal evaluations, then compared their evaluations with a partner or group members, and settled on a numeric assessment for the metric. Monitors did this for all 13 RBP measures, then totaled them into an overall score.  As well as observing professional monitors, the first author had to chance to experience their learning process firsthand; they made RBP assessments alongside other monitors, learned through group discussions, and performed assessments as part of a group (figure ~\ref{fig:ExpertAssessment}).

\subsubsection{Volunteer Training and Data Collection}
\begin{figure}[th]
\centering
  \includegraphics[width=1\columnwidth]{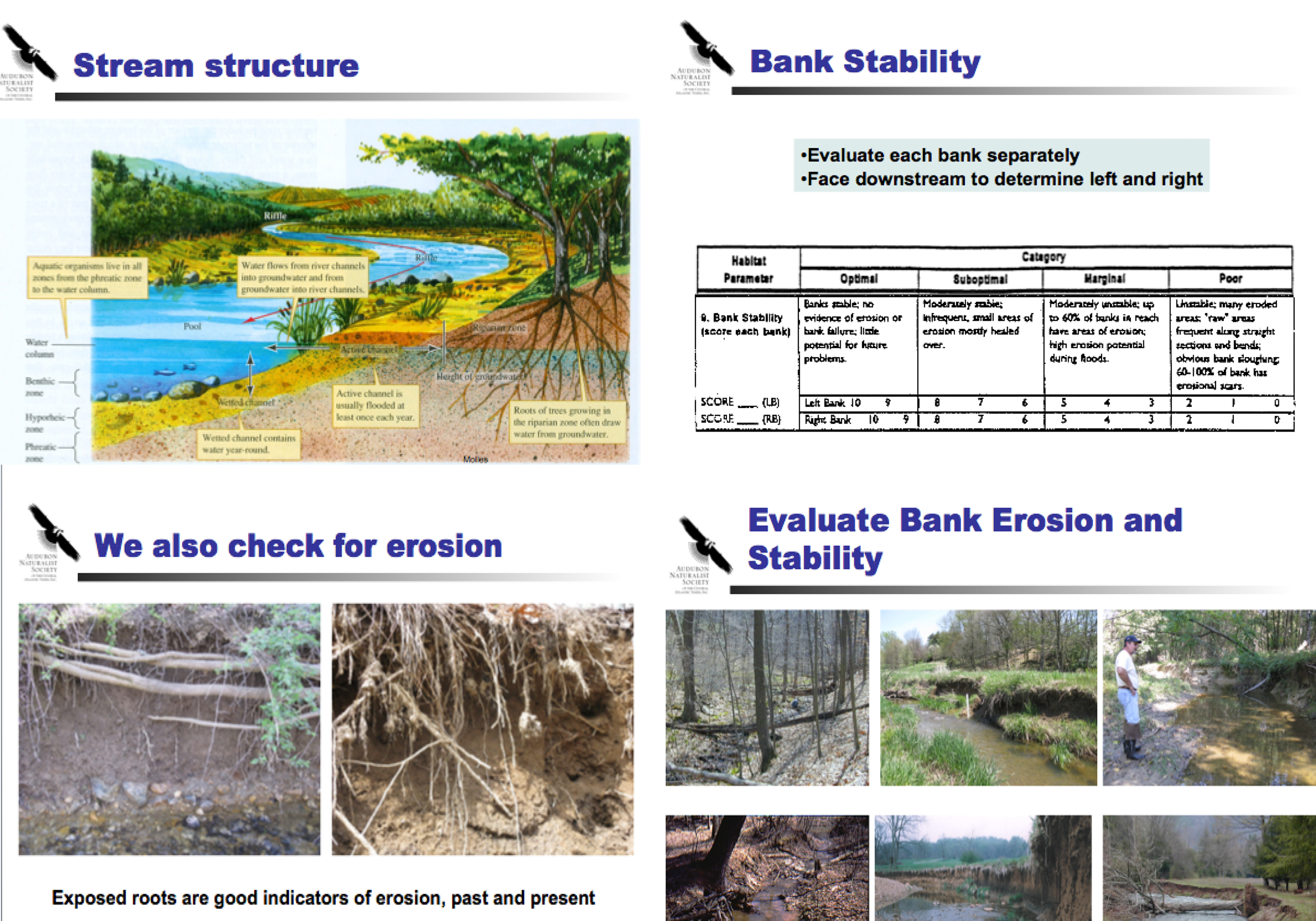}\caption{Examples of the RBP training slides. The top left slide overviews background information about the stream, the top right slide describes the ``Bank Stability'' protocol, and the bottom slides show images of streams with different bank stabilities.}~\label{fig:CS_Slides}
\end{figure}

The first author also observed and participated in volunteer RBP training\footnote{Although volunteers do not make qualitative assessments during data collection, they are introduced to the RBP scales as a background to water monitoring.}. Unlike professionals, who learn onsite, monitors were introduced to the protocol through a 3 hour PowerPoint lecture~\cite{Wiss:ReadStream} with images that exhibited a range of quality for the RBP stream characteristics (figure ~\ref{fig:CS_Slides}).

\begin{figure}[th]
\centering
  \includegraphics[width=1\columnwidth]{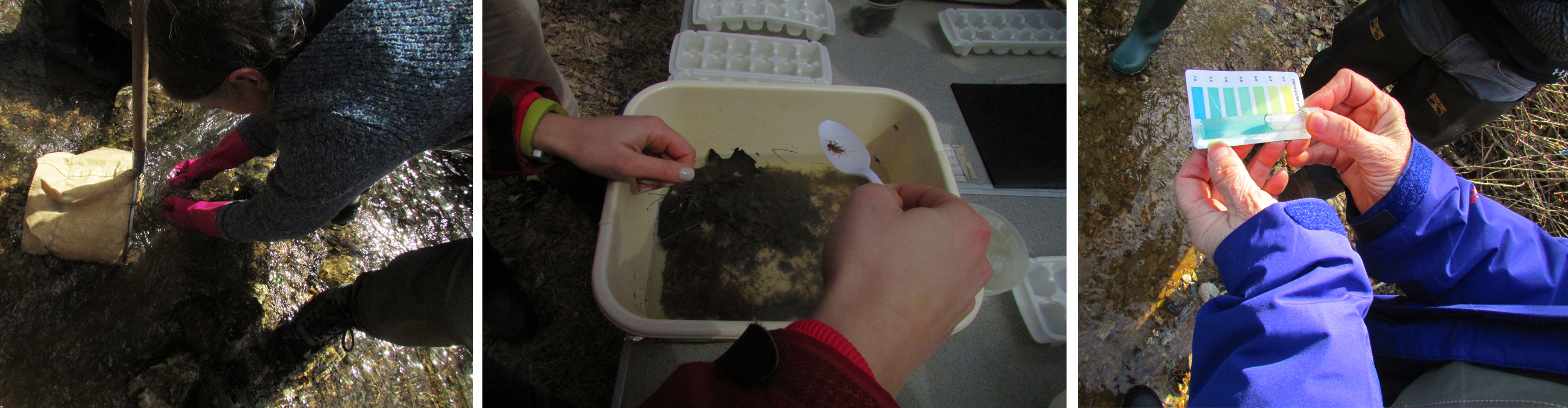}\caption{Images from the volunteer outdoor data collection. During data collection, volunteers collected and counted macroinvertebrates, measured PH and temperature.}~\label{fig:CS_DataCollection}
\end{figure}

As well as experiencing PowerPoint training, the first author participated in a 3 hour outdoor training and data collection experience (figure~\ref{fig:CS_DataCollection}) with 10 volunteer monitors. During data collection, monitors collected macroinvertebrates with hand-held fishing nets, separated them by species using a field guide, and counted them. After categorizing and tallying each species, volunteers practiced measuring the stream's PH and temperature.

\subsection{Informal Discussions with Ecologists}
We received initial insights about the qualitative stream assessment process at the Association of Mid-Atlantic Aquatic Biologists (AMAAB), a regional conference for water biologists and ecologists. During the conference, the first author presented a poster describing early findings based on the in-person training experiences. During the 2 hour poster session, the author discussed her findings with more than 15 ecologist, biologists, and water monitors. They also collected feedback on how the RBP protocol was employed by different monitoring groups, how her first-person training experiences compared to researchers' own training, and whether volunteers could be trained virtually to perform these tasks.

\subsection{Semi-Structured Interviews with Professional RBP monitors}
We conducted 5 in-depth interviews with professional monitors by phone or in person (after on-site training), each of which lasted approximately 1.5 hours. Participants included an aquatic biologist at the EPA, a program manager of the the Virginia Department of Environmental Protection, and an ecologist at the Fairfax Department of Public Works and Environmental Services. Interviewees had between 2 and 23 years of experience making qualitative stream assessments, including one participant who helped develop and test the protocol in 1992. During each interview, participants described (1) their personal process for making RBP assessments, (2) how they learned to make assessments, and (3) how they taught young professionals to make assessments. In addition, we conducted an interview with the water quality monitoring program coordinator at at the Audubon Naturalist Society~\cite{Audubon_mathias} to understand the nature of volunteer water monitoring tasks and training.

Several interview and feedback sessions were audio-recorded, however recording was not  possible for all sessions, either due to a loud outdoor environment or a conference setting.  When recording was not possible, the first author took detailed notes and followed up with participants for further discussion and clarification.We then listened to recordings, and transcribed important quotes. We chose not to transcribe recordings in their entirety because of budget and time constraints. However, we accounted for this by taking detailed notes during the interviews, intermittently stopping interviewees to repeat important words or phrases.  
\subsection{Data Analysis}
We compiled notes, annotations, and quotes from the training and data collection experiences, informal discussions, and expert interviews into a single spreadsheet, and thematically open-coded important themes~\cite{strauss1990basics}. Our open-coding process took several stages of iteration. First we split the notes, annotations, and quotes into individual statements. Then, we color-coded related statements, grouped them into categories, and ascribed a theme name to each category (this process was similar to performing thematic analysis by grouping color-coded post-it notes~\cite{braun2006_thematicAnalysis}). We then iterated on the categories by regrouping statements until we felt that they fit together cohesively.  As we reorganized the statements, we renamed the category themes to most closely fit the patterns we observed. Once we were satisfied with our precise themes, we grouped related themes together, and identified a final set of themes and sub-themes. 

%% file: Results_Process.tex
The following section compares current teaching methods employed by citizen science groups and professional water monitors, and describe challenges of making qualitative stream assessments. Then, we describe salient themes from expert interviews and feedback sessions. 

\subsection{Differences in Background}
We found that the professional monitors we met had a more extensive background in the natural sciences than the volunteers.  All the professional monitors we interviewed either had a degree or had taken multiple courses in biology, ecology or conservation before focusing on water monitoring. In addition, professionals mentioned completing water-quality accreditations or certificates to learn to perform procedural tasks like fish and macroinvertebrate sampling. Although they had related training, the professionals we spoke to did not have a specific background in RBP assessment, they learned to do the qualitative assessment through their first-person experiences.

In contrast to professionals, volunteers that participated in water monitoring training and data collection had a range of background experiences: some had degrees in biology or ecology, others had participated in other monitoring training, and a few had little to no experience in the domain, but were eager to learn. Notably, almost all of the volunteer monitors we spoke to during the citizen science outdoor training session had some sort of higher education background: several of the younger volunteers were recently out of college, and many of the older volunteers were retired educators. We do not have exact data on volunteer backgrounds because we learned about them informally, through conversations during outdoor data collection. However, finding that volunteers were highly educated is supported by previous work suggesting that citizen science attracts affluent volunteers motivated to improve society~\cite{raddick2009galaxy}. 
			
In addition, the Audubon society coordinator discussed several background courses that helped support volunteers with different backgrounds. Volunteers could learn about ecology through courses like  \textit{a natural history of aquatic ecology} and \textit{healthy stream biology}. Likewise, volunteers could build procedural and identification skills through classes such as an \textit{overview of invasive plants} and a series on \textit{aquatic insect identification}~\cite{Audubon_mathias}. Volunteers could even take certification exams to participate in certain projects or lead volunteer teams. 

While professional monitors had more formal training than volunteers, a surprising finding is that there was less of a distinction between professional and volunteer monitoring backgrounds than we anticipated. Both groups had a higher education, had some experience in the natural sciences, and had opportunities to become better monitors through certifications. Notably, the biggest difference between professionals and volunteers was the number of streams each group had experience evaluating; professionals visited many more streams than volunteers, and were thus able to more easily compare them.

Since professional monitors had the opportunity to first-hand experience many more streams than volunteers, the way they learned about the qualitative RBP measures differed greatly from volunteers. Professionals learned to make assessments in an apprenticeship under more experienced monitors, whereas volunteers learned about the measures through a PowerPoint lecture that outlined the protocol measures, but did not transfer any nuanced or practical assessment skills.   
\subsection{Challenges Interpreting Qualitative Measures}
As well as finding differences in background and learning between professional and volunteer monitors, our research uncovered multiple challenges in interpreting the protocol. The task of interpreting a protocol is quite similar in nature to interpreting a survey question; in order to make an informed response, a participant has to understand the meaning of the words in the question, understand the scale dimensions, and know how to map the question to the scale~\cite{fowler1995improving}. Similarly, in order to make an informed RBP assessment, an evaluator has to understand the contextual definitions of quality defined by each protocol metric, then consider what quality attributes correspond to scales values.

A primary challenge of the RBP is characterizing the state of an outdoor environment using a quasi-quantitative scale. To make an informed assessment, monitors must make subjective interpretations of multiple scale descriptions that are hard to quantify. Table~\ref{tab:RBP_Challenges} illustrates several interpretation issues that exist, including \textit{interpreting scale measures}, \textit{accounting for site variability,} \textit{evaluating related measures}, and interpreting and \textit{accounting for the passage of time}. 

For instance, during a professional training session, monitors explained that a scale measuring embeddedness of particles in a stream bed (RBP protocol 2a~\cite{barbour1999rapid}) should not be interpreted linearly, even though it was written to suggest linearity: the written scale suggests  that 25\% is suboptimal quality and 75\% is poor quality. Professional monitors explained that realistically, more than 25\% embeddedness should be characterized as poor quality because the environment becomes unsuitable for macroinvertebrates. Given this discrepancy, it is would not be clear to a volunteer monitor whether to evaluate  25\% embeddedness as suboptimal or poor.

\begin{figure}
  \includegraphics[width=1\columnwidth]{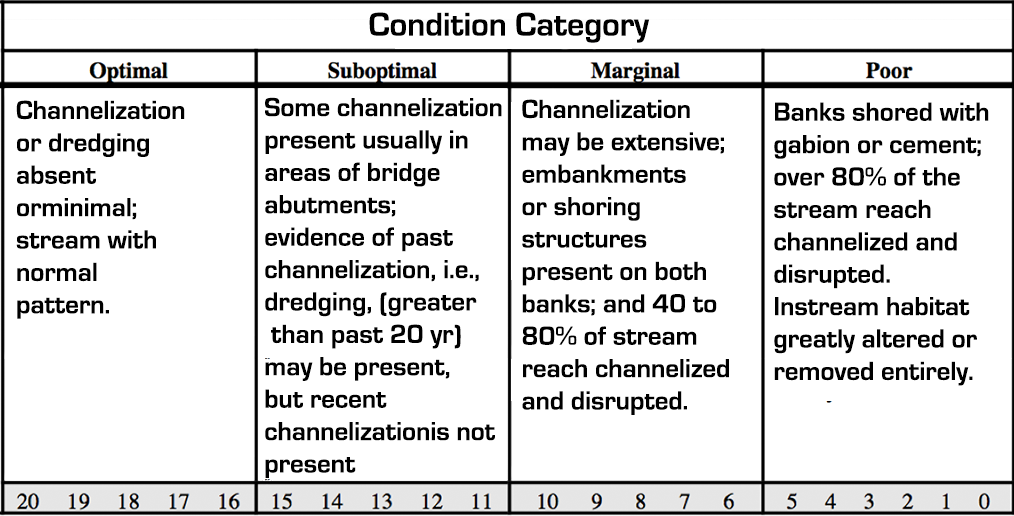}
    \caption{\textit{Channel Alteration}, a qualitative EPA metric that evaluates human impact on stream channels. This metric poses several challenges, including asking monitors to interpret the meaning of ``\textit{normal pattern}'' streams and subjective time metrics (e.g.``g\textit{reater than the past 20 years}''), and to judge percentages of channel disturbance.}~\label{fig:ChannelAlteration}
\end{figure}

Similarly, some protocol measures describe quality using time measures that require application of heuristics. For example, in channel alteration (RBP protocol 6, shown in figure~\ref{fig:ChannelAlteration}, the suboptimal condition asks monitors to evaluate if there is evidence of channelization (stream straightening, widening or deepening) ``greater than past 20 years.'' This assumes that monitors can heuristically estimate the timeframe of a disturbance. Further, this time measure is used asymmetrically, only to describe the suboptimal category. This increases the challenge of making evaluations because it is difficult to compare the suboptimal category to other categories. 

Our experiences found that the subjectivity and imprecision of the RBP protocol compelled monitors to rely background knowledge and personal experiences to make assessments. Notably, our informal conversations with ecologists revealed that the water monitoring community was aware of interpretation challenges that existed in the protocol and made up for the protocol flaws with personal contextual knowledge.

Although they heavily relied on personal experiences, professionals suggested that their divergent monitoring experiences biased their assessment. For instance, an ecologist in Fairfax, Virginia worked with primarily urban disturbed streams, whereas a West Virginia monitor evaluated undisturbed rural streams.  Due to their different backgrounds, the Fairfax ecologist was more likely to judge stream quality more leniently than the West Virginia monitor. 

In line with these discrepancies, Roth~\cite{roth2002biological} remarks that the monitoring community must resolve issues in field sampling protocols, differences in types of data collected, and discrepancies in condition ratings. Further, Roth appeals for increased accuracy of stream condition estimates.

\begin{table}
\centering
\begin{tabular}{L{2.5cm}|L{10.5cm}}
\hline
\textit{Protocol Interpretation Challenges} & Example \\
\hline
\textit{How to interpret scales?} & The RBP protocol suggests that measures should be assessed linearly, but professionals suggest linearity varies between measures.  For example, professionals interpret a stream with 25\% or more embeddedness as poor because the environment is unsuitable for macro-invertebrate organisms even though the protocol considers the habitat suboptimal.  \\
\hline
\textit{How to Account for Site Variability?} & Data collectors are asked to evaluate 100 meter stream cross sections, but how should users evaluate areas containing with significant variability? Professionals make holistic judgments of quality based using their experience, but new data collectors have no foundation with which to make such judgments. \\
\hline
\textit{How to evaluate related measures?} & Several measures of stream quality directly affect one another (e.g. stream bank stability affects sediment deposits). How should data collectors account for this in their assessments? \\
\hline
\textit{How to interpret passage of time?} & 3 of 10 protocol measures ask users to evaluate transience of stream elements (e.g. logs and cobble) and recency of human activity (e.g. whether stream channel alteration occurred more or less than 20 years ago). How would users know how to judge the passage of time? \\ 
\hline
\end{tabular}
\caption{\label{tab:RBP_Challenges}}
\end{table}

%% file: Results_Themes.tex
\subsection{Expert Interview Findings}
As well as revealing the nature of professional monitoring experiences, the expert interviews uncovered multiple themes that illuminate professionals' qualitative assessment process. These themes help us understand how monitors make observations in outdoor stream environments, how they compare streams, identify patterns, and use these patterns to make RBP assessments.

\subsubsection{Making Intuitive Judgments of Quality}
Our expert interviews revealed that professional monitors had a complex relationship with the RBP protocol. Professionals suggested that together, the 13 measures helped capture ``a snapshot'' of stream quality, but many agreed that the individual measures were either imprecise and challenging to interpret. This supports the first author’s experience learning to make RBP assessments with professional monitors, and parallels our informal discussions with professionals.

We found that professionals with different amounts of expertise make RBP assessments very differently; less experienced monitors dutifully tried to interpret the protocol language, whereas more experienced professionals developed an intuition for assessment quality.  Two interviewees, a senior biologist and ecologist, said they linked ``mental images'' from their experiences to the protocol scales. Likewise, the program manager who helped develop the protocol suggested that after making evaluations for 24 years, they could evaluate a stream at a glance, ``without even scoring it.'' While the RBP scales are technically quantitative, expert interviews confirmed our theory that that practically, the measures are qualitative. 
\subsubsection{Using Multisensory Information to Make Judgments}
The professionals we interviewed and the ecologists we spoke to at the AMAAB conference described using multisensory environmental information to form opinions of a stream's quality. Several biologists mentioned supplementing the EPA protocol with additional measures for trash, presence of human activity, and invasive plant and animal species. For instance, an ecologist commented that hearing European Starlings (an invasive species) was indicative of poor stream quality. Likewise, another noted that invasive plant species often indicated that a stream habitat has probably been disturbed. Although these heuristics were not formally part of the qualitative assessment protocol, professionals suggested that paying attention to these characteristics helped them locate environment stressors that affected the RBP measures.

Professional monitors also used ambient sensory information to take note of habitat stressors  during observations. For example, one ecologist approximated the strength of stream riffles by the sound of rushing water, whereas another used sun warmth and wind strength to judge the density of stream bank vegetation. Since formal assessment was limited to 100 meter cross-sections, monitors conferred that additional sensory information allowed them to make more comprehensive observations about the state of the stream that they may not have readily been able to see. This is in-line with Dede’s 1999 findings~\cite{dede1999multisensory} that multisensory information helps students understand complex scientific models through experiential metaphors and analogies. This is likewise in line with Dede’s (2017)~\cite{dede2017virtual} work on stream identification tasks, finding that sensory information (\textit{e.g.} sound, color and turbidity of the water, weather variables, shifts in grass color) could help learners sense pattern changes.

\subsubsection{Describing Stream Quality using Past, Present, and Future Narratives}
 As well as observing sensory information and making intuitive judgments, several monitors we spoke to actively interpreted `narratives' of the stream spaces: how the landscape had transformed from the past, and how the stream in its present state would shape its future.  For instance, during an on-site learning experience at a stream in northern Virginia, a professional monitor pointed to markers indicating that a stream was part of a historical agriculture site. They explained that the former use of the land had caused extensive erosion at the monitoring site by irrigating water through the stream, causing faster moving water that wore away at the stream’s banks. Using the stream’s history and current state, the professional then predicted that in the next  5-10 years the stream would become wider and more eroded.

Likewise, during an expert interview, a biologist emphasized how connected the ecosystem was, and described the stream landscape as a consequence of multiple events. They recounted that sedimentation was often causes by people mowing a stream’s vegetative zone in order to keep snakes off of their property. Then they explained that removing bank vegetation removes the roots holding soil in place, and that the next time a stream floods, the whole bank erodes.
Rather than merely evaluating streams in their current forms, professional monitors predicted what caused the stream characteristics, and how those stream characteristics would change over time. 

%% file: Discussion.tex
\section{Discussion}
The goal of this work was to assess the viability of training citizen science volunteers to make qualitative assessments of streams and watersheds using the EPA’s Rapid Bioassessment Protocol (RBP)~\cite{barbour1999rapid}. We aimed to do this by examining the challenges of interpreting the RBP, and discerning between professionals and volunteer water monitor experiences. 

Our findings suggest that the biggest difference between professional and volunteer monitors are volunteers' lack of experience with on-site streams; professionals experience many different streams throughout their career, whereas volunteers only see a handful of streams, likely in their immediate neighborhood or community.

Although we did not formally interview volunteers about their experience evaluating streams, our findings suggest that professionals use their vast experience to make intuitive judgments of streams that extend beyond RBP measures: they make intuitive judgments of quality, interpret stream characteristics using multisensory information, and are able to describe past, present, and future narratives of different streams. That professional monitors make intuitive judgments can be at least partially explained by the challenges of making qualitative assessments using the RBP: the protocol asks professionals to subjectively interpret differences in quality based on their experiences and to account for misleading measures.

\subsection{Volunteer Training Needs}
Professionals develop intuitive judgments by visiting many streams of differing quality. In order for volunteers to be able to skillfully interpret the RBP, they too would need to experience first-hand a range of diverse streams. Further, our findings suggest that professionals scaffold their intuitive knowledge through environmental sensory information. Volunteers may also need to experience this environmental information to develop similar intuitive judgments of quality. This confirms findings from Psotka~\cite{psotka1995immersive} and Dihn~\cite{dinh1999evaluating}, who suggest that multisensory cues can reduce conceptual load, create salient memories and emotional experiences, and increase memory and sense of presence for environmental information.

Our findings also suggest that the process of discussing and reviewing intuitive judgments with peers helps professionals review and update their internal information scaffold, which is in line with Crossan's work on creating shared interpretations of meaning~\cite{crossan1999organizational}. Effective training should thus include some sort of assessment loop, allowing learners to discuss and receive feedback from peers or teachers. 

Finally, our work found that rather than making isolated assessments, many professionals interpret the state of a stream as part of a changing narrative. This suggests that professionals develop higher-order thinking skills, inductively envisioning the stream's history and future based on their knowledge of how stream features affect one another. This finding strongly echoes the literature; climate scientists use qualitative heuristics to monitor change over time, whereas paleoecologists use qualitative heuristics to reconstruct the past. This is also in line with Bloom's education taxonomy~\cite{krathwohl2002revision_BloomTaxonomy}, which suggests that  beyond analyzing and evaluating information, students should be able to reassemble information into new ideas. To become effective monitors, learners must be trained to consider  how environmental features together impact stream characteristics over time.  

\subsection{Training Solutions}
Our findings suggest that volunteers need to experience streams the way that professionals do, and to learn from peers and experts. However, citizen science training is often constrained by time and monetary resources~\cite{bonney2009citizen,wiggins2013free}; personally training a crowd of volunteers at several streams is infeasible and unrealistic.  Recent advances in {360\textdegree} panoramic videos, augmented reality (AR), and virtual reality (VR) have made it possible to remotely give learners the experience of evaluating different streams, and getting feedback from peers or experts. While it is not in the scope of this paper to compare their value, the following sections describe each technology, and table~\ref{tab:Tech_Affordances} summarizes the affordances and drawbacks of each.

\begin{table}
\centering
\begin{tabular}{L{2cm}|L{5.5cm}|L{5.5cm}}
\hline
 Technology & Affordances & Drawbacks \\
\hline
\textit{{360\textdegree} Panoramic Videos }
& 
\begin{itemize}
\setlength\itemsep{1em}
\item Flexible viewing modalities
\item Cheap to view on an HMD
\item Real video footage
\end{itemize}
& 
\begin{itemize}
\setlength\itemsep{1em}
\item Minimal interaction 
\item Expensive to produce  
\item Requires careful video calibration 
\end{itemize} 
 \\
\hline
\textit{Augmented Reality (AR)} 
& 
\begin{itemize}
\setlength\itemsep{1em}
\item Interact with the real world 
\item Maintain environmental awareness
\end{itemize}
& 
\begin{itemize}
\setlength\itemsep{1em}
\item Imprecise visual rendering 
\item Expensive to develop
\item Expensive to deploy
\item Poor support for multi-user experiences 
\end{itemize} 
\\
\hline
\textit{Virtual reality (VR)} 
& 
\begin{itemize}
\setlength\itemsep{1em}
\item Sense of presence in virtual worlds
\item Interact with many environments in same location
\end{itemize}
& 
\begin{itemize}
\setlength\itemsep{1em}
\item Can cause simulator sickness
\item Expensive to develop 
\end{itemize} 
\\
\hline
\end{tabular}
\caption{\label{tab:Tech_Affordances}}
\end{table}

\subsubsection{Panoramic Videos}
{360\textdegree} videos are spherical video recordings that record every direction around a camera. Viewers can playback video either on a flat display by controlling panoramic viewing direction, or by projecting the panorama onto a series of projectors or in a head mounted display (HMD)~\cite{garza_2015_360}. {360\textdegree} videos are easy to view on any computer or cheap smartphone HMD such as Google Cardboard. Likewise, {360\textdegree} videos are realistic, since they capture actual stream footage. Although easy to view, high quality videos can be expensive to produce, since they require either a special rig of multiple cameras or a dedicated camera that simultaneously films overlapping angles. Further, to create the {360\textdegree} effect, footage must be carefully stitched and calibrated~\cite{butler2011_vermeer_360}. In addition, the videos only allow users to change the panorama view, but do not support support interaction with the environment.

\subsubsection{Augmented Reality (AR)}
AR is a direct or indirect live view of a real-world environment ``augmented'' by computer-generated perceptual information. Overlaid information can be constructive (\textit{i.e.} additive to the natural environment) or destructive (masking of the environment) and is spatially perceived as an aspect of the real environment~\cite{van2010_ARsurvey}. Although spatially realistic, AR is expensive to develop because development kits are still in their infancy. Further, AR often struggles with imprecise visual rendering, and does not support multi-user experiences~\cite{van2010_ARsurvey}.

\subsubsection{Virtual Reality (VR)}
VR is a computer-generated scenario that simulates physical experiences~\cite{UltimateVR_TechnologyGuide}. The immersive environment can be similar to the real world or it can be fantastical, creating an experience not possible in our physical reality. VR allows users to look around, move within, and interact with an artificial world. Current technologies pair the VR experiences with realistic images, sounds, and sensations. VR's sense of presence in a virtual world allows users to interact with different environments, features, and people in any location~\cite{dalgarno2010_VRlearningAffordance}, however, development is expensive and time-consuming. Further, may cause participants to develop temporary simulator sickness~\cite{kennedy1993_simulatorsickness_Questionnaire}.

The technologies described offer distinct benefits for qualitative training, but also suffer from distinct challenges. For instance, {360\textdegree} video offer high realism and cheap deployment, but is not interactive. In contrast, AR offers realism and interactivity, but suffers from lack of precision. Relative to {360\textdegree} videos and AR, VR creates the greatest sense of presence and interaction in any number of virtual worlds, but is not photo-realistic. 

\subsection{Future Work}

Future work should consider which tools most easily allow learners to flexibly and cheaply experience a range of environments. To do this, we plan to rigorously weigh the affordances and drawbacks of {360\textdegree} videos, AR, and VR against the volunteer training needs we identified earlier in the discussion, and identify whether one technology is most effective for training.

In order to select the most appropriate technology, we must consider how to implement training that addresses the high-level volunteer needs we identified. For instance, we found that volunteers need to experience different stream environments, but it is not clear what kind of realism they need in order to compare their observations and synthesize them into patterns. Literature suggests that multisensory information improves learning and judgment tasks~\cite{dematte2007olfactory,brooks1990project,lee_Spence2008_MultimodalDriverFeedback,yannier2016adding}, but it is not clear how much realism is necessary in training. Is photorealistic training necessary to scaffold experiences into an effective cognitive map? Likewise, Is multisensory realism necessary?

Similarly,  future work must consider how virtual training should support learner discussion and negotiation of meaning. Should training include discussion with peer learners, with professionals, or with both? Can peers or professionals be simulated with non-player characters (NPC's), or must they be real people? If they are real, should peer learners or professionals interact symmetrically or asymmetrically with one another?  To support discussion and negotiation of meaning, future work must also consider how to support joint action and coordination, which are built on the spatiotemporal coherency of a shared interaction space~\cite{knoblich2011psychological}.

Finally, training design must consider how to help learners unite their stream quality observations with background knowledge to form stream quality narratives. Which tools support learners from a range of backgrounds? How might training be designed to help learners reassemble their stream observations into a temporal narrative?

%% file: Conclusion.tex
\section{Conclusion}

The overarching goal of our work was to use the water monitoring domain to assess the viability of training volunteers to make different types of inductive qualitative assessments. To undertake this challenge, we used the domain of water monitoring to understand how professionals train and make qualitative assessments, and how volunteers learn about qualitative assessments. To do this, we observed and participated in RBP training and data collection with professional water monitors and with with citizen scientists, informally discussed water monitoring methods with ecologists, and conducted semi-structured interviews with professional monitors.

We found that professionals learned to make assessments on-site, through iterative assessments and discussions with peers and instructors. From their experience, we found that professionals develop intuitive judgments of quality, use multisensory environmental information to make judgments, and construct past and future narratives of the stream using environmental characteristics.  We also found that the qualitative RBP protocol is subjective and misleading, perhaps because it tries to characterize intrinsically qualitative measures.

Contrary to our expectations, we found that volunteers primarily differed from professionals in the number of streams they had visited and assessed. To match professional training experiences, we identified 3 training needs; to first-hand experience environmental information in order to develop intuitive judgments, to discuss judgments with other monitors, and to form quality narratives from assessments.
\subsection{Generalizing to other Domains} 
While our work focused on the domain of water monitoring, our findings inform other ecology citizen science projects like eBird, IceWatch, and the Clean Air Coalition, which assess physical, ecological, and societal variables~\cite{mckinley2017_CitizenScienceUse_Examples}. A goal of these projects is to help the public understand and appreciate complex ecosystems, and to engage them in the task of identifying problems and solutions. Training citizen scientists to interpret qualitative measures related to these projects could help support this goal. For instance, birdwatchers could learn to assess ecosystem habitats based on the presence of different types of birds. Likewise, ice and pollution monitors could learn to identify relationships between observable and climate factors. Paired with quantitative measures, learning to make qualitative assessments can help volunteers understand and appreciate complex relationships that comprise their world.